\documentclass[12pt,preprint]{aastex}

\newcommand{\Rsun}{\ensuremath{R_{\sun}}}
\newcommand{\rside}{\ensuremath{r_{\rm side}}}

\newcommand{\Msun}{\ensuremath{M_{\sun}}}

\newcommand{\K}[1]{\hbox{$K_{\rm #1}$}}
\newcommand{\M}[1]{\hbox{$M_{\rm #1}$}}
\newcommand{\R}[1]{\hbox{$R_{\rm #1}$}}
\newcommand{\Teff}{\hbox{$T_{\rm eff}$}}
\newcommand{\geff}{\hbox{$g_{\rm eff}$}}
\newcommand{\vsini}{\ensuremath{v\hspace{0.005in}\sin\hspace{0.005in}i}}
\newcommand{\kmps}{km~s$^{-1}$}

\newcommand{\chisq}{$\chi^{\rm 2}$}

\newcommand{\Na}[1]{\ion{Na}{#1}}
\newcommand{\Fe}[1]{\ion{Fe}{#1}}

\newcommand{\pubdao}{Publ.\ Dominion Astrophys.\ Obs.\ Victoria}

\shorttitle{The Classical $\zeta$~Aurigae Binaries}
\shortauthors{Eaton, Henry, \& Odell}

\begin{document}

\title{Orbits and Pulsations of the Classical $\zeta$ Aurigae Binaries}

\author{Joel A.\ Eaton\altaffilmark{1}, 
        Gregory W.\ Henry\altaffilmark{1}, and 
        Andrew P. Odell\altaffilmark{2}}

\altaffiltext{1}{Center of Excellence in Information Systems,
       Tennessee State University, Nashville, TN; 
       eaton@donne.tsuniv.edu}
\altaffiltext{2}{Northern Arizona University, Flagstaff, AZ}

\begin{abstract}
We have derived new orbits for $\zeta$ Aur, 32~Cyg, and 31~Cyg with observations 
from the TSU Automatic Spectroscopic Telescope and used them to identify non-orbital 
velocities of the cool supergiant components of these systems.  We measure periods 
in those deviations, identify unexpected long-period changes in the radial velocities, 
and place upper limits on the rotation of these stars.  These radial-velocity variations 
are not obviously consistent with radial pulsation theory, given what we know about the 
masses and sizes of the components.  Our concurrent photometry detected the {\it non-radial} 
pulsations driven by tides (ellipsoidal variation) in both $\zeta$~Aur and 32~Cyg, at a 
level and phasing roughly consistent with simple theory to first order, although they 
seem to require moderately large gravity darkening.  However, the K component of 32~Cyg 
must be considerably bigger than expected, or have larger gravity darkening than 
$\zeta$~Aur, to fit its amplitude.  However, again there is precious little evidence 
for the normal radial pulsation of cool stars in our photometry.  H$\alpha$ shows some 
evidence for chromospheric heating by the B component in both $\zeta$ Aur and 32~Cyg, 
and the three stars show among them a meager $\sim$ 2--3 outbursts in their winds of 
the sort seen occasionally in cool supergiants.  We point out two fundamental questions 
in the interpretation of these stars, (1) whether it is appropriate to model the surface 
brightness as gravity darkening and (2) whether much of the non-orbital velocity structure 
may actually represent changes in the convective flows in the stars' atmospheres.
\end{abstract}


\section{Introduction}

Our detailed knowledge of stars in the main sequence comes from analyses of 
eclipsing double-lined spectroscopic binaries.  Solutions to light and velocity 
curves of such objects can define masses and radii of the component stars well 
enough to challenge the details of calculated internal structure and evolution.  
In contrast, defining the basic properties of evolved stars is normally much more 
difficult.  The long periods and correspondingly large separations of binaries 
containing them make eclipses unlikely, and the existing binaries tend to be 
only single-lined.  The $\zeta$~Aur binaries, however, with their eclipses and 
composite spectra, give us a unique opportunity to determine physical properties 
of a few massive supergiant stars reliably in the same way we can for many 
main-sequence stars.  A good example of this is the way Bennett et al.\ (1996) 
defined the properties of $\zeta$~Aur.  Wright (1970) discussed these systems, 
particularly the three classical systems $\zeta$~Aur, 31~Cyg, and 32~Cyg, all 
three of which have supergiant K primaries paired with B stars close to the main 
sequence.  Table 1 gives their fundamental properties.

Most close binaries have circular orbits, although all possible eccentricities 
seem equally likely among newborn systems, at least those with longer periods 
(e.g., Abt 2006).  The three classical $\zeta$~Aur systems all have sizable 
eccentricities.  For this reason, they ought to be subject to certain binary 
proximity effects in ways other stars are not.  For instance, they will be subject 
to a non-radial pulsation driven by the variable tidal distortion inevitable in an 
eccentric binary (Cowling 1941; Eaton 2008; Sepinsky et al.\ 2007).  This phenomenon 
is equivalent to the ellipsoidal variation from the equilibrium tidal distortion 
in circular, synchronously rotating binaries.  Because of the different orientations 
of their orbits, the two closer systems $\zeta$~Aur and 32~Cyg would manifest 
different phase dependence of this effect in ways giving clues about the internal 
properties of supergiants.  Guinan \& McCook (1979) claimed to have detected this 
phenomenon in 32~Cyg.  Wilson (1979) included a theory for it in the Wilson--Devinney 
code for calculating binary light curves.  The other major proximity effect, the 
so-called reflection effect, might be detectable in these systems as well.

All the cool giants and supergiants seem to be variable, probably through radial 
pulsations.  Henry et al.\ (2000) argued that all stars to the red of the 
Linsky--Haisch coronal dividing line are pulsational variables.  Even the red giants 
to the blue of it are variable given precise radial velocities (Walker et al.\ 1989).
The components of $\zeta$~Aur binaries would be expected to manifest the pulsations 
of similar supergiants.  Such pulsations would be in addition to the aforementioned 
proximity effects.  Differences in their pulsational periods {\it might} give us an 
idea of how mass is distributed in their interiors.

Zeta Aur itself is the most interesting of the three classical systems in terms of 
its binary interactions.  It has the shortest period (970 d) and biggest eccentricity 
($e$ $\sim$ 0.4), and these qualities make it most interesting for looking for the 
effects of a tidally driven non-radial pulsation.  Griffin (2005) discussed the orbit 
recently, using all the many radial velocities then available.  Why, then, should 
we waste our time redoing his analysis?  Our data are several times as precise, cover 
one orbit continuously, and thereby begin to show coherent deviations of the K star 
from its orbital velocity. 

We will (1) improve the orbital elements for two of the three classical $\zeta$~Aur 
binaries, (2) assess the rotation of these stars, (3) model the ellipsoidal light 
variations in order to interpret the driven pulsations of the cooler components of 
these systems,  (4) look for the intrinsic (radial) pulsations of these stars and 
use them to restrict the radii, and (5) look for evidence of proximity effects 
and other variation in H$\alpha$.

\section{Observations}

Observations consist of new spectra and photometry for the three classical systems, 
spanning roughly 3.5 years since 2004.  All these data come from the completely 
automatic observatory Tennessee State University (TSU) maintains at Fairborn 
Observatory, a private site in southern Arizona (Eaton, Boyd, \& Henry 1996).

\subsection{Spectra}

We observed $\zeta$~Aur, 31~Cyg, and 32~Cyg between JD\,2,452,860 and JD\,2,454,200, 
obtaining echelle spectra of roughly 30,000 resolution, with the TSU 2-m Automatic 
Spectroscopic Telescope (Eaton \& Williamson 2004, 2007).  This set consists of 302, 
217, and 348 useful spectra, respectively, for the three stars.  We reduced and 
analyzed them with standard pipeline techniques to derive radial velocities and 
equivalent widths of H$\alpha$.  These measurements are available electronically as 
Table 2.  Listed are (1) HJD, the Heliocentric Julian Date of observation (minus 
2,400,000), (2) $RV_{\rm cool}$, the radial velocity of the K star, (3) EW1, an 
equivalent width of H$\alpha$ absorption, and (4) EW2, an equivalent width of enhanced 
absorption in the blue wing of H$\alpha$.  The other column is a tag identifying 
the star by its HD number.  Missing data in this table are identified with a ``9.999".

The measured velocities from the AST have a formal external error of 0.10--0.11 \kmps\ 
and are 0.35$\pm$0.09 \kmps\ more negative than the IAU radial-velocity system (Eaton \& 
Williamson 2007).  The velocities in Table 2 are the raw velocities without the +0.35 
\kmps\ correction to the IAU system.  Values of systemic velocities, $\gamma$, from our 
orbital solutions, listed in Table 4, are transformed to the IAU system.

Our H$\alpha$ data consist of observed equivalent widths in a wide band (6561.3--6565.05~\AA\ 
in the rest frame of the star), EW1, designed to measure the total absorption in the normal 
profile of such a star, and a narrow band (6559.7--6561.3~\AA), EW2, to detect enhancements 
of the blue wing of the profile that may signal episodic mass ejections in a star's wind.  
In measuring the spectra, we adjusted the continuum to a common level by defining 13 
pseudocontinuum points in the range 6522--6600~\AA, automatically measured their levels in 
the spectra, and renormalizing the spectra to line segments between those points.  The EW's 
depend on the two points at 6559.3 and 6568.1~\AA.  Several hundred spectra of the K giants 
$\alpha$~Tau, $\alpha$~Boo, and $\alpha$~Ari, which ought to be $\sim$ constant in H$\alpha$, 
give standard deviations per measurement of 0.046 and 0.020 \AA, respectively, for EW1 and EW2.  
We shall use these values as the uncertainties of measurement.

\subsection{Photometry}

We also collected $BV$ observations of the three stars with the TSU 0.4-m Automatic Photometric 
Telescope (APT), obtaining measurements over complete cycles of both $\zeta$~Aur and 32~Cyg 
and 4.0 years for 31~Cyg.  These measurements consist of nightly means of differential 
measurements with respect to a comparison star, HD~34412 for $\zeta$~Aur and HD~192985 for 
both 31 and 32~Cyg.  The check star for $\zeta$~Aur was HD~30834, with 32~Cyg observed as a 
check star for 31~Cyg.  These data should have an external error near 0.004 mag (Henry 1995).  
They are available electronically as Table 3.  Data listed are HJD$-$2,400,000, 
($\Delta$$U$,$\Delta$$B$,$\Delta$$V$)$_{\rm variable}$, 
($\Delta$$U$,$\Delta$$B$,$\Delta$$V$)$_{\rm check}$--when available, and HD number of the 
star.  We identify missing data, such as the nonexistent $\Delta$$U$'s, with a ``99.999".
This arrangement preserves the format of photometric data available on our Internet site.

\section{Analysis}

We shall analyze the three stars to find out how their radial velocities deviate from 
purely orbital motion and combine these results with photometry to assess what forms 
the pulsations and proximity effects take in them.

\subsection{Deviations from Orbital Velocities}

The great precision of our AST data lets us solve velocity curves of these three 
long-period binaries and look for deviations from elliptical orbits.  Table 4 gives the 
results for the three stars, listing the usual spectroscopic elements.  For 31~Cyg we have 
only about one-third of a full orbital cycle of data, so we have combined our data with 
those listed by Wright \& Huffman (1968) and weighted all the data equally.  The solution 
to this combined data set is the same as Wright \& Huffman's to within the putative errors.  
The difference between the spectroscopic period we have derived and the photometric period 
(1.8~d) corresponds to a shift of $+$0.21 \kmps\ of Wright \& Huffman's velocities with 
respect to ours.  This shift gives a flavor of the kind of uncertainties that indeterminate 
zero-point shifts introduce into orbital analyses.  The values of the major elements ($K$, 
$e$, and $\omega$) for both $\zeta$~Aur and 32~Cyg agree with previous determinations to 
within the likely errors of those determinations.  In particular, they agree with Griffin's 
(2005) values for $\zeta$~Aur to within 2$\sigma$ of his formal errors.  For 32~Cyg, they 
should be a significant improvement on the elements of Wright (1970), with which they 
agree to within the likely errors of that analysis.  This excellent agreement means that 
the shapes and orientations of the orbits of both stars are known well enough to support 
rigorous analyses of their atmospheric eclipses, driven nonradial pulsations, and eclipse 
timings. 

Figure 1 shows the velocity curves of the three stars.  All three are obviously variable 
with non-orbital shifts superimposed on the dominant orbital velocities.  Figure 2 shows 
the time dependence of these deviations, which can be rather extreme.  The 250 d, 0.75 
\kmps, deviation of 31~Cyg around 53700, for instance, if pulsational, would correspond 
to a 23 \Rsun\ change in the stellar radius, about 12\%, even without any allowance for 
foreshortening.  Alternatively, it could represent some sort of truly global circulation.  
The variation seen in Figure 2 seems cyclic on timescales of 100--300 d, so one might 
suspect  that some of it could be seasonal observational effects.  However, the major 
effects do not correlate very well in 31 and 32~Cyg, which we observed over the same 
observing season, occasionally on the same nights, and there is absolutely no hint of such 
effects in velocities of HD~14214 at levels above 0.05 \kmps\ (Eaton \& Williamson 2007).  
Three other K supergiants observed over roughly the same time interval, $\epsilon$~Peg 
(K2~Ib), $\xi$~Cyg (K4--5 Ib-II), and 63~Cyg (K4 Ib-IIa), show long-term variations at 
least as great as the three binaries, although we have far fewer data for these single 
stars.

We have used two techniques to look for periodicity in the residuals.  In the first, using 
a program written by D.\ S.\ Hall, we fit sine curves for a spectrum of periods to the data 
[$\Delta$$RV$ = $A$\hspace{0.015in}$\sin$(2$\pi\hspace{0.005in}$HJD/$P$ + $\phi$)] and 
identified minima of \chisq\ of these fits as possible periods.  In the second, we applied 
the techniques of Van\'i\^cek (1971), as we have for the multiperiodic $\gamma$~Doradus stars 
(e.g., Kaye et al.\ 1999; Henry et al.\ 2001).  This second approach lets us reliably find 
multiple periodicites without ``prewhitening," an advantage, especially in the low-frequency 
domain.  We searched for periods in the range 1--1200~d.  Both methods identified essentially 
the same periods, but, because the second method gives more systematic results, we will use 
them in the following analysis.  Table 5 lists the periods found and amplitudes of sinusoids 
fit to the data for them.  If the velocity variation represents a radial pulsation, we may 
integrate the (sinusoidal) variation over half a cycle to get the total excursion in radius, 
$\Delta$$R$ = $\xi$$AP$/$\pi$, where $\xi$ $\approx$ 1.35 is a correction for the fact that 
expansion of most of the disc is only partially in the line of sight (e.g., Gray \& 
Stevenson 2007).  The periods derived here likely reflect the timescales of some physical 
phenomena in these stars but not truely coherent long-duration pulsations.  This is especially 
so for the longest periods, those comparable to the $\sim$1200~d duration of our observations. 
Additional tests for shorter periods (0.03--1.0~d) with the method of Van\'i\^cek found none, 
as expected.


\subsection{Rotation}

If the cool components of these binaries were rotating synchronously, they would have 
significant rotational velocities, \vsini\ = (\K1+\K2)\R1/a for synchronous rotation with 
the usual assumptions about orientation of the motions.  For pseudosynchronous rotation 
(Hut 1981; Hall 1986), the velocity would be even bigger.  For $\zeta$~Aur, the values 
would be 8.3 and 16.6 \kmps, respectively, with the enhancement for pseudosynchronous 
rotation calculated with Hut's eq.\ 42.  Rotational velocities, hence line broadening, 
for the other two systems would be less, an unobservable \vsini\ = 2.71 and 3.6 \kmps\ 
for 31~Cyg, for example, although it should be observable for 32~Cyg at \vsini\ = 6.50 
and 10.4 (or 10 and 15 if we adopt the much bigger relative radius implied by tidal 
distortion).  If $\zeta$~Aur were rotating pseudosynchronously, we could easily detect 
it by comparing profiles of its metallic lines with those in 31~Cyg and other, single, 
stars.  A couple of well exposed single spectra of $\zeta$~Aur and 31~Cyg do not show 
shallower, hence broader, lines in $\zeta$~Aur than in 31~Cyg, nor do the spectra of 
$\zeta$~Aur binaries and other K supergiants plotted by Eaton (1995, Fig.\ A19) show 
an apparent difference in depths of strong metallic lines.  To get an idea of the 
magnitude of the expected broadening, we artificially broadened a spectrum of the 
K2~Ib supergiant $\epsilon$~Peg to \vsini\ = 8.3 and 16.6 \kmps\ and compared the line 
profiles in the broadened spectra with unbroadened profiles.  Both values gave measurably 
shallower lines, by 10 and 30\%, respectively.


To quantify this result, we have looked at the strengths and depths of strong metallic 
lines in three other cool supergiants, $\epsilon$~Peg, $\xi$~Cyg, and 63~Cyg, plotting 
them up with composite spectra for the three $\zeta$~Aur binaries for phases when the 
K supergiants were not illuminated by their B companions.  Depths of strong \Fe1 lines 
in the three single stars vary by $\sim$ 2\%.  The lines in $\zeta$~Aur 
itself in these composites were actually somewhat shallower than in most of the other stars, 
but by only of the order of 3\%.  This implies a rotational velocity $\sim$ 30\% synchronous, 
or 2.5 \kmps.  On the other hand, 32~Cyg, which ought to be rotating about 
as fast as $\zeta$~Aur, has lines as deep as in any of the single stars.

Another way to gauge the rotation in these systems is to look for shifts of shell lines 
formed in the lower chromosphere during the atmospheric eclipses.  Many investigators 
have done this.  Griffin et al.\ (1990) argued they had found a displacement 8.5 \kmps, 
corresponding to synchronous rotation, in a single precise observation of $\zeta$~Aur.  
Earlier measurements of such displacements in $\zeta$~Aur were not so clear cut.  McKellar 
\& Petrie (1952) found that metallic lines in a 1950 eclipse gave displacements of only 
2.5 \kmps, possibly from rotation much slower than synchronous.  Wilson \& Abt (1954) 
found positive shifts in ingress and negative shifts in egress, as expected for rotation, 
but these shifts came with significant scatter and a temporal dependence different from 
that of rotation.  McKellar \& Butkov (1956) found that lines of ionized metals in the 
near ultraviolet, which would be unblended with photospheric lines, followed the velocity 
of the K star to a few \kmps\ in 1955--1956.  Bauer (1994) reviewed the evidence for 
31~Cyg, finding that velocities of the metallic shell lines (e.g., \Fe1) were always 
about the same as the orbital velocity and that in most cases the differences were not 
consistent with rotation.  Because it has a grazing total eclipse, 32~Cyg might not be 
expected to show rotational displacements of metallic shell lines, and it does not 
(Wright \& Hesse 1969).

These stars are clearly not rotating pseudosynchronously, or even synchronously {\it 
unless the single stars are rotating much faster than generally thought}, and 
we see no convincing evidence they are rotating significantly faster than single stars.
In conducting this analysis, we have implicitly assumed that single K supergiants are 
not rotating, inasmuch as we have no theoretically calculated comparison spectra for 
such stars, nor would we trust them if we did.  Analyses of bright giants (class II stars)
by Gray \& Toner (1986a) find rotation \vsini $\lesssim$ 3 \kmps, although Gray \& Toner 
(1987) find higher rotation (\vsini $\lesssim$ 7 \kmps) for the class Ib supergiants.
Eventually it ought to be possible to directly test the idea that K supergiants are 
rotating at rates like those inferred by Gray \& Toner by using more precisely measured 
velocities of chromospheric shell lines in these binaries.

\subsection{The Non-Radial Pulsation from Ellipsoidal Light Variation}

Guinan \& McCook (1979) analyzed the light outside eclipse for 32~Cyg and concluded they 
had found the ellipsoidal variation of that star.  Fredrick (1960, Fig.\ 4) had previously 
detected variation looking like ellipsoidal variation on the even longer period of VV~Cep.

Figure 3 shows the measured brightnesses of the three stars over the past 4.0 years in 
the $V$ band; data for $B$ are similar.  We seem to have detected the ellipsoidal variation 
of both $\zeta$~Aur and 32~Cyg, along with its strong periastron effect around primary eclipse.  
The phase dependence for 32~Cyg is not as clear cut as in $\zeta$~Aur or in Guinan \& McCook's 
photometry, probably because of intrinsic variation. However, these two sets of photometry 
for 32~Cyg actually agree fairly well, especially close to periastron.  Our data for 32~Cyg 
for the last two seasons illustrated have unexpected shifts that mask the periastron effect 
to some extent.  The light curve for 31~Cyg is incomplete, although it does show a sinusoidal 
wave (P = 1200~d) over our period of observation that is not consistent with ellipsoidal 
variation.  If this wave actually results from variation of their common comparison star, 
correcting for it would make the brightness of 32~Cyg more consistent with the calculated 
ellipsoidal variation. 

The theoretical curves plotted in Figure 3 represent the ellipsoidal variation for these 
binaries calculated with the Wilson--Devinney code roughly for the elements given in 
Tables 1 and 4.  They assume the stars are hardly rotating (0.1 synchronous) and incorporate 
the usual assumptions about the surface brightness, namely linear limb darkening ($x$=0.80) 
and Lucy's (1967) convective gravity darkening.  However, it is not clear that a star with a 
driven non-radial pulsation could legitimately have its surface-brightness distribution 
specified by gravity darkening.  Gravity darkening is an equilibrium, diffusive theory that 
supposes timescales much longer than the typical pulsation, driven or intrinsic.  Furthermore, 
we know that pulsating stars can have a complicated, double-valued dependence of temperature 
on radius.  If we assume to first order that {\it radially} pulsating stars are both apparently 
brighter and hotter when they are smaller, we might parameterize their variation of surface 
brightness, $F$, with gravity, $\nabla\Omega$$\sim$$r^{-2}$, as gravity darkening 
[$F_{\rm Bol}$ $\propto$ $\nabla\Omega^{g}$ $\propto$ $r^{-2g}$].  Then the gravity exponent, 
$g$, would be $\sim$ 1.0 for such a star to stay the same brightness as it pulsates.  Adiabatic 
pulsation would give a much higher effective value of $g$.  Intrinsically driven pulsations of 
stars are far from adiabatic, and this leads to phase lags between the radial compression and 
the star's brightness, which depend on details of the star's structure and driving mechanism 
(e.g., Szab\'o, Buchler, \& Bartee 2007).  Non-radial pulsators are theoretically more 
complicated because horizontal motions relieve horizontal pressure variations, at least for 
non-radial pulsation much slower than the natural frequency of the star.  Buta \& Smith (1979) 
calculated the light variations of non-radial pulsators, finding that these horizontal 
adjustments greatly reduce the temperature variations predicted by Dziembowski's (1971) theory 
for adiabatic pulsation.  In fact, they found that geometrical effects alone can explain most 
of the light variations of some actual non-radially pulsating stars.  Furthermore, their 
calculations implied that the temperature variation might well be period dependent.  What 
phase lags one might expect of non-radial pulsation is also open to conjecture (e.g.\ 
Townsend 2003).  Theory obviously does not give us especially good guidance for predicting 
how temperature varies over the surface of a star with a driven pulsation.

We may get a better idea of how flux depends on gravity as a star pulsates by considering 
some actual pulsating stars.  As a first stab at doing this, we have looked at analyses of 
$\beta$~Cep stars, which are often suspected of having nonradial pulsations (e.g., Stamford 
\& Watson 1977; Odell 1980).  Calculations used to predict the complicated line profiles 
for assumed pulsation modes (e.g., Stamford \& Watson 1976) often do not even include the 
effect of pulsation-induced effective temperature variations.  However, we can get an estimate 
of that effect in an actual star by using Kubiak's (1972) analysis of BW~Vul as an example.
The values of effective temperature and effective gravity in his Table 6, admittedly somewhat 
double valued, give a slope of $\Delta$log{\Teff}/$\Delta$log{\geff} $\approx$ 0.18, or 
g $\approx$ 0.7.  This is larger than one might expect form extant theories of gravity 
darkening, but even contact components of Algol binaries, which seem to conform to the 
assumptions behind the theory, seem to require $g$ $\sim$ 0.5, somewhat larger than Lucy's 
0.32 (Eaton 2008).  

The data for $\zeta$~Aur in Figure 3 and geometry derived by Bennett et al.\ (1995) are 
precise enough to give a crude test of the theory incorporated into the Wilson-Devinney code.
The solid curve is calculated for \rside=0.163\footnote{This \rside\ is the radius in the 
orbital plane perpendicular to the line between the centers of the two stars, measured in 
units of the orbital separation (semimajor axis).} (148 \Rsun) with $g$=0.32.  The dashed curve 
represents the case \rside=0.163 with $g$=1.0.  The calculated phase dependence is excellent, 
meaning that there is no appreciable phase lag in the non-radial pulsation.  The calculated 
amplitude is too small for $g$=0.32 but somewhat large for $g$=1.0.  From this evidence, it 
appears the effective gravity darkening is $g$ $\approx$ 0.7--0.8.  The theory thus seems 
to work to first order and restricts the radius to better than about 10\%, as we judge for 
other calculations for $R$ = 130 and 182 \Rsun.  

The theory fails to predict the ellipsoidal variation of 32~Cyg, however, and this failure 
implies the radius must be much bigger than thought, as Guinan \& McCook found in their own 
analysis, or the gravity darkening must be {\it much} larger than allowed by $\zeta$~Aur.  
We have plotted the predicted variation for two cases in Figure 3, $R$=217 \Rsun\ (dashed) and 
$R$=271 \Rsun\ (solid), both for $g$ = 0.7.  The radius required to fit the amplitude with 
$g$=0.7 is \rside=0.23$\pm$0.02 (260$\pm$20\Rsun).  The phase dependence of the periastron 
effect might be improved with a bigger eccentricity.  However, the orbital elements are 
known very reliably from the radial velocities, and we therefore see no justification for 
using $e$ as an aphysical fitting parameter in a light-curve solution.

Our rather crude analysis of the tidal distortion is about all anyone could expect to 
do without a much better theory of the surface brightness of these stars with driven 
pulsations.  

\subsection{Intrinsic Variation (Pulsational?)}

Cool giants as a group seem to be pulsating, and the supergiant components of $\zeta$~Aur 
binaries should be no exception.  Our new photometry has very little evidence of pulsation 
beyond that driven by tides, but the radial velocities plotted in Figure 2 show unmistakable 
evidence of cyclic, if not periodic, variation in all three stars (see Table 5).

As for the photometry, it is difficult to detect long-term periods in any of these stars 
because of the strong, somewhat poorly modelled ellipsoidal light variation.  There is 
obvious deviation from the calculated light curves in Figure 3.  Thirty-two Cyg seems to 
have a coherent sinusoidal pulse of 105~d period around 53125, which may have been a 
random pulsation.  There seems to have been an anticorrelated change in the radial 
velocity, but the data sets did not overlap very well for that year, and they were rather 
noisy at the level of this effect.  There also seems to be quasiperiodic variation of the 
brightness on shorter periods (e.g., near 53500) but at the level of the noise in these 
data.

We may estimate the pulsational periods expected for these stars by using theoretical 
pulsational constants (e.g., Fox \& Wood 1982) [$Q = P(M/\Msun)^{1/2}(R/\Rsun)^{-3/2} 
\propto  P\sqrt{\,\overline{\rho}}$~] and mean densities derived from the information 
in Table 1.  All three stars have roughly the same mean density, which is about 
one-tenth that of $\alpha$~Tau.  We would thus expect longer pulsational periods, by 
several times, than found in the normal giants.  Values are $Q$/$P$ = 1.34, 1.23, and 
1.35 $\times$ 10$^{-3}$, respectively, for $\zeta$~Aur, and 31 and 32~Cyg.  The expected 
fundamental periods ($Q$$\sim$0.08--0.18) would be near 100 d, and the first overtone 
($Q$$\sim$0.03--0.04), near 25 d.  Alternatively, we can invert this process and calculate 
$Q$'s for observed periods to get the values listed in the fifth column of Table 5.  These
values are much larger than expected from any known pulsational mechanism, and are 
reminiscent of the unexplained long-period variations observed in many AGB stars (Wood, 
Olivier, \& Kawaler 2004; Hinkle et al.\ 2002).  

Whether the periodic variations we see in the radial velocities represent pulsation or some 
other phenomenon is open to conjecture.  Given the lack of {\it photometric} variation 
correlated with these variations of the radial velocities, we think it unlikely that the 
velocities result from pulsation of the stars.  The aforementioned 105~d pulse of 32~Cyg has 
the right length for the radial fundamental, especially if we accept the reality of the 
apparent pulses at periods closer to the first overtone.  This pulse has an amplitude 
$\sim$~0.5 \kmps\ peak-to-peak, giving an excursion of only 1.9~\Rsun, or about 0.7--1.1\% 
of the radius of the star.  Likewise, the apparent 125~d periodicity in the radial velocities 
of 31~Cyg, if a radial pulsation, implies that the radius changes by 0.7~\Rsun, or 0.4\%.  
Some of the longer periods detected in radial velocities, however, imply much bigger changes in 
radius, the 429 d periodicity of 32~Cyg (0.4 \kmps\ peak-to-peak), for example, corresponding 
to 4.4~\Rsun.  Furthermore, the singular excursion in the velocity of 31~Cyg near 53700 
corresponds to a change of $>$12\% of the radius, if pulsational.

If the changes in radius implied by cyclic radial velocities reflect radial pulsation, they 
must affect the timing of eclipses, and there is some evidence of this effect in the 
literature.  When the radius of the star changes by 1--2\%, the duration of total 
eclipse must also change by a comparable amount, roughly 0.5~d in systems like $\zeta$~Aur 
and 31~Cyg, so eclipse timings would give a way of testing for the existence of pulsations.  
The problem here is that eclipse timing is very difficult to measure in these systems, 
and the typical photometric analysis is compromised by being pieced together from data 
from many sources for the same eclipse.  The best results seem to be for 31~Cyg, for 
which Stencel et al.\ (1984) determined times very precisely for the eclipses of 1962 
and 1982.  We can use their precise ephemeris to predict times in the 2004 eclipse 
observed with the APT and place a limit on how much the timing varies.  Their $t_{50}$ 
for ingress of 1982, extrapolated forward by 2 cycles, agrees with our observations to 
at least $\pm$0.1~d.  This is a $\pm$0.3\% change in the semiduration, or $\sim$ 0.6 
\Rsun\ in radius.

\subsection{The Reflection Effect and Behavior of H$\alpha$}

A reflection effect in these stars could take several forms, (1) heating of the 
atmosphere, detectable in changes of temperature-sensitive line ratios with phase, 
(2) increased ionization of the photosphere, detectable in weakening of very strong 
lines like \Na1 D and strengthening of lines of singly ionized species, and (3) 
higher ionization in the chromosphere, detectable as a change in H$\alpha$--probably 
an increase in its strength.

To look for effects of direct heating of the atmosphere, we have made composites of 
spectra of $\zeta$~Aur and 32~Cyg at phases when the B star is behind the K star and 
when it is in front.  If there are any differences in the photospheric absorption 
lines, they are very subtle.  We looked for enhancements of such potentially 
sensitive lines as \Fe2 $\lambda$6416.93 and 6432.68, finding a possible {\it very} 
slight enhancement with the hot star in front.  The \Na1 D lines may have been a bit 
weaker in the irradiated spectrum of 32~Cyg, and H$\alpha$ was definitely stronger 
in both stars, as seen in Figure 4.  

Direct heating of the chromosphere by the B star could potentially change the strength
of H$\alpha$.  There are at least two ways H$\alpha$ might vary in cool supergiants, 
(1) an overall change in the mass of the $\sim$ hydrostatic chromosphere where the bulk 
of the line forms (e.g., Cram \& Mullan 1979) and (2) an enhancement of the blue wing 
in the stellar wind (e.g., Mallik 1993), which is observed occasionally in supergiants
(e.g., Smith \& Dupree 1988; Eaton \& Henry 1996).

Figure 4 shows the time/phase dependence of H$\alpha$ in these three stars.  We see 
variations much greater than the expected observational errors or the variations 
expected in normal (class III) giants. The few single supergiants and bright giants
measured by Eaton (1995) showed a much greater range of equivalent width, however,
and the values for our three $\zeta$~Aur components fall into that range.  Furthermore, 
the three supergiants $\epsilon$~Peg, $\xi$~Cyg, and 63~Cyg have fluctuations of this 
amount during the same span of time.  The phase dependence of H$\alpha$ in the 
three binaries does not correlate neatly with their orbits.  The strong enhancement 
seen in 32~Cyg over the range 53200--53550, for instance, occurred roughly when the 
cool star's irradiated face was pointing towards us, but its superior conjunction 
was near the end of this range, and the range does not coincide at all well with the 
time between the ascending and descending nodes, marked in the figure.

Our new data for these stars show very few of the eruptions/enhancements of winds seen 
in other supergiants.  There are 4 instances of enhanced absorption in the blue wing 
of H$\alpha$ apparent in Figure 5.  An enhancement for $\zeta$~Aur near 53840 seems 
to be a real change in the stellar profile.  The elevation for 31~Cyg near 53800 is 
also probably stellar, as well.  However, the high points for 32~Cyg marked ``?" in the 
figure are probably telluric, and the high points for 31~Cyg near 54000 could be also.  
On the other hand, the general enhancement in the absorption after 53500 for 31~Cyg seems 
to reflect a genuine change in the profile.

\section{Discussion and Summary}

The photometry and radial velocities of these three stars raise more questions about 
pulsation than they answer.  In only one very restricted case do we see something like 
a pulsation in both the brightness and radial velocity of the star.  This pulse in 32~Cyg 
may have been a radial pulsation of the K star, and the same star shows flickering 
on shorter timescales, at the level of the photometric errors, that may be pulsation 
in overtones.  The roughly coherent variation of 31~Cyg's radial velocity at 125~d 
may also reflect pulsation, but it is not accompanied by changes in the brightness.
Radial pulsation at the 1--2\% level is consistent with eclipse timings in these stars.

The K supergiants in these systems fall in a part of the HR diagram with rather low 
pulsation (e.g., Maeder 1980; Henry et al.\ 2000).  Stars with lower masses are generally 
stable in their lower radial modes but become increasingly susceptable to pulsation in high 
overtones (Xiong \& Deng 2007).  Such high overtones are a possible source of the apparently 
random flickering of the rather stable K giants and supergiants.

One way to get a better idea of the level of any changes in radius from pulsation is to 
look critically at the timing and duration of eclipses, as we have illustrated with three 
eclipses of 31~Cyg.  We do not think the existing data are good enough to do this in any 
meaningful way.  However, this approach should be possible with photometry from robotic 
telescopes, but it would take a communal effort over many years. 

Other changes in both brightness and velocity are completely inconsistent with the known 
stability of light variation of these systems and with the expected pulsations of their 
K components.  Especially perplexing is the 200-day drop in velocity of 31~Cyg, 
corresponding to a 12\% change in the star's radius.  This kind of change would be 
accompanied by changes of several days in the eclipse timing.  It is much more likely 
to be a nonpulsational change in the circulation of the star's atmosphere, like the famous 
starpatch in $\xi$~Boo~A (Toner \& Gray 1988).  The range of photospheric velocity 
caused by granulation or other flows in cooler supergiants (Gray \& Toner 1985, 1986a, 
1986b; 1987) seems big enough ($\sim$ 6--10 \kmps) to admit fluctuations at the level we 
are observing.  However, once again we are thus reminded that ``there are more things 
in heaven and earth than are dreamt of in [our] philosophy."

All three of these stars had photometric variations on timescales longer than expected 
for pulsation.  Thirty-one Cyg showed a 1250~d sinusoidal variation in our photometry, 
covering $\sim$ the length of our observations.  There is no mechanism for producing 
this effect, and it may simply reflect variation of the comparison star.  Both $\zeta$~Aur 
and 32~Cyg also had variations in brightness beyond their ellipsoidal variation.

Cool components in these three classical systems seem to be rotating no faster than similar 
single supergiants.  In contrast, Griffin et al.\ (1993) and Eaton \& Shaw (2007) found 
evidence the chromosphere of 22~Vul is rotating {\it faster} than synchronously.  This is 
a close binary in a circular orbit, which may have been a much closer, interacting system 
in a previous visit to the giant branch.  Likewise, the supergiant component of the 
relatively close but eccentric binary HR~6902 (G9~II+B8~V) seems to be rotating even 
faster than pseudosynchronously (Griffin \& Griffin 1986).  These rotational velocities 
would seem to be an important clue to the evolutionary history of supergiant binaries 
once somebody becomes clever enough to interpret them.

We have detected the ellipsoidal variation and its periastron effect in two of the stars 
and used it to discuss how the driven nonradial pulsations in such a star should change 
the surface brightness.  In this context, we question the use of the concept of gravity 
darkening in such stars and propose a methodology for determining an effective gravity 
darkening for such pulsations.  We find the light variations of $\zeta$~Aur require 
larger gravity darkening than predicted by Lucy's (1967) diffusive theory.
Along these same lines, we may have detected a chromospheric reflection effect 
in the H$\alpha$ strength.

\acknowledgments

This research used the SIMBAD database and was supported by NASA grants NCCW-0085 and NCC5-511
and by NSF grants HRD~9550561 and HRD~9706268.

{\it Facilities:} \facility{TSU:AST}\


\begin{deluxetable}{cccrrrrrrl}
\rotate
\tablecaption{The Classical $\zeta$ Aur Systems}
\tablewidth{0pt}
\tablehead{
\colhead{Star} & \colhead{Spectrum} & \colhead{Period} & \colhead{$a$}    & 
\colhead{$i$}  & \colhead{\R1}      & \colhead{$r$}    & \colhead{\M1}    & 
\colhead{\M2}  & \colhead{References}\\
\colhead{}     & \colhead{}         & \colhead{d}        & \colhead{(\Rsun)}& 
\colhead{(deg)}  & \colhead{(\Rsun)}& \colhead{(\R1/$a$)}& \colhead{(\Msun)}&
\colhead{(\Msun)}& \colhead{}        \\
\colhead{(1)} & \colhead{(2)} & \colhead{(3)}    & \colhead{(4)} &
\colhead{(5)} & \colhead{(6)} & \colhead{(7)}    & \colhead{(8)} &
\colhead{(9)}& \colhead{(10)}
}
\startdata
 $\zeta$~Aur   & K4~Ib + B5~V      &   972&   905&  87.0&  148&  0.163&   5.8&  4.8& 
                                                               Bennett et al.\ (1996) \\
    HD~32068   &                   &      &      &      &     &       &      &     &  \\
      31~Cyg   & K4~Ib + B3--4     &  3784&  2710&  87.2&  197&  0.073&  11.7&  7.1& 
                                                               Eaton (1993b);         \\
   HD~192577   &                   &      &      &      &     &       &      &     & 
                                                               Eaton \& Bell 1994)    \\
      32~Cyg   & K4--5~Ib+ B6--7   &  1148&  1130&  78.6&  175&  0.155&   9.7&  4.8&
                                                               Eaton (1993a) \\
   HD~192909   &                   &      &      &      &     &       &      &     &  \\
\enddata
\tablecomments{Spectral types are from Wright (1970).  Other quantities are from the 
cited references in col.\ 10}
\end{deluxetable}


\begin{deluxetable}{rccrrrrc}
\tablecaption{Sample Spectroscopic Data}
\tablewidth{0pt}
\tablehead{
\colhead{HJD}          & \colhead{$RV_{\rm cool}$} & \colhead{EW1}   & \colhead{EW2}  & 
\colhead{Star} & \\
\colhead{(2,400,000+)} & \colhead{(\kmps)}         & \colhead{(\AA)} & \colhead{(\AA)}&
\colhead{}     & \\
\colhead{(1)}          & \colhead{(2)}             & \colhead{(3)}   & \colhead{(4)}  &
\colhead{(5)}  &
}
\startdata
 52861.9477&  0.11&  1.334&  0.113& HD~32068& \\
 52863.8894&  0.02&  1.496&  0.152& HD~32068& \\
 52895.9664&  3.60&  1.426&  0.167& HD~32068& \\
 52896.9640&  3.70&  1.407&  0.149& HD~32068& \\
 52897.9219&  3.87&  1.446&  0.161& HD~32068& \\
 52898.9425&  3.94&  1.447&  0.162& HD~32068& \\
 52899.9319&  3.93&  1.419&  0.154& HD~32068& \\
\enddata
\tablecomments{This excerpt shows the form and content of the full table 
as published in the electronic edition of the Astrophysical Journal.}
\end{deluxetable}


\begin{deluxetable}{cccccccc}
\tablecaption{Sample Photometric Data}
\tablewidth{0pt}
\tablehead{
\colhead{HJD}& 
\colhead{$\Delta$$U_{\rm var}$}& \colhead{$\Delta$$B_{\rm var}$}& \colhead{$\Delta$$V_{\rm var}$}&
\colhead{$\Delta$$U_{\rm chk}$}& \colhead{$\Delta$$B_{\rm chk}$}& \colhead{$\Delta$$V_{\rm chk}$}&
\colhead{Star} \\
\colhead{(2,400,000+)}& \colhead{}& \colhead{}& \colhead{}& \colhead{}& \colhead{}& \colhead{}& 
\colhead{} \\
\colhead{(1)}& \colhead{(2)}& \colhead{(3)}& \colhead{(4)}& \colhead{(5)}& \colhead{(6)}&
\colhead{(7)}& \colhead{(8)}
}
\startdata
 52895.9881& 99.999& $-$0.327& $-$0.942& 99.999&  0.886&  0.085&  HD 32068 \\
 52926.0065& 99.999& $-$0.324& $-$0.934& 99.999&  0.874&  0.079&  HD 32068 \\
 52930.0235& 99.999& $-$0.322& $-$0.933& 99.999&  0.876&  0.081&  HD 32068 \\
 52931.0199& 99.999& $-$0.324& $-$0.939& 99.999&  0.877&  0.077&  HD 32068 \\
 52932.0135& 99.999& $-$0.327& $-$0.939& 99.999&  0.878&  0.079&  HD 32068 \\
 52933.0132& 99.999& $-$0.327& $-$0.940& 99.999&  0.876&  0.080&  HD 32068 \\
 52934.0214& 99.999& $-$0.326& $-$0.938& 99.999&  0.871&  0.069&  HD 32068 \\
\enddata
\tablecomments{All of this table is published in the electronic edition 
of the Astrophysical Journal.  This excerpt shows its form and content.}
\end{deluxetable}


\begin{deluxetable}{rclllcrcl}
\tabletypesize{\scriptsize}
\rotate
\tablecaption{Spectroscopic Orbits}
\tablewidth{0pt}
\tablehead{
\colhead{HD} & \colhead{Period} & \colhead{T\tablenotemark{a}} & \colhead{K}       & 
\colhead{$\gamma$\tablenotemark{b}} & \colhead{e} & \colhead{$\omega$} & \colhead{References} \\
\colhead{}   & \colhead{(days)} & \colhead{HJD$-$2,400,000}    & \colhead{(\kmps)} & 
\colhead{(\kmps)}                   & \colhead{}  & \colhead{(deg)}    &  \colhead{}           \\ 
\colhead{(1)}& \colhead{(2)}    & \colhead{(3)}                & \colhead{(4)}     & 
\colhead{(5)}                      & \colhead{(6)}& \colhead{(7)}      & \colhead{(8)}
}
\startdata
  $\zeta$ Aur    &  (972.162)& 53039.9$\pm$0.10   &  23.17$\pm$0.02 &  \ 10.81$\pm$0.01& 0.3973$\pm$0.0007& 328.9$\pm$0.13& This paper \\
  31~Cyg         & 3786.1$\pm$0.7& 52372.8$\pm$2.2&  13.78$\pm$0.13 &  $-$7.41$\pm$0.08& 0.224$\pm$0.006  & 206.4$\pm$1.4 & This paper \\
  32~Cyg         &  (1147.80)& 53796.9$\pm$0.28   &  16.64$\pm$0.03 &  $-$7.45$\pm$0.02& 0.3126$\pm$0.0014& 222.1$\pm$0.3 & This paper \\
\enddata
\tablenotetext{a}{Periastron passage.} \tablenotetext{b}{Velocity on IAU system.}
\tablecomments{Values in parentheses are assumed values taken from the literature.  
                               Periods are generally from Batten et al.\ (1989).}
\end{deluxetable}

\clearpage


\begin{deluxetable}{lcccc}
\tablecaption{Possible Periods in Velocity Residuals}
\tablewidth{0pt}
\tablehead{
\colhead{Star} & \colhead{Period}& \colhead{$A$}    & \colhead{$\Delta$$R$} & \colhead{$Q$}\\
\colhead{}     & \colhead{(d)}   & \colhead{(\kmps)}& \colhead{(\Rsun)}     & \colhead{(d)}\\
\colhead{(1)}  & \colhead{(2)}   & \colhead{(3)}    & \colhead{(4)}         & \colhead{(5)}
}
\startdata
$\zeta$~Aur&  252.3 $\pm$ 2.4 & 0.111 $\pm$ 0.011 &  1.5 & 0.34 \\
           &  431.6 $\pm$ 5.0 & 0.131 $\pm$ 0.011 &  3.0 & 0.58 \\
           &  658.8 $\pm$ 10.2& 0.078 $\pm$ 0.010 &  2.7 & 0.88 \\
           &  186.8 $\pm$ 0.9 & 0.063 $\pm$ 0.009 &  0.6 & 0.25 \\
31~Cyg     & 1062 $\pm$ 26    & 0.287 $\pm$ 0.016 & 16~~ & 1.31 \\
           &  542 $\pm$ 10    & 0.108 $\pm$ 0.012 &  3.1 & 0.67 \\
           &  125.2 $\pm$ 0.5 & 0.081 $\pm$ 0.009 &  0.5 & 0.15 \\
           &  347.2 $\pm$ 3.0 & 0.050 $\pm$ 0.009 &  0.9 & 0.43 \\
           &  166.4 $\pm$ 0.8 & 0.063 $\pm$ 0.009 &  0.6 & 0.20 \\
           &  201.1 $\pm$ 1.0 & 0.045 $\pm$ 0.009 &  0.5 & 0.25 \\
32~Cyg     &  721 $\pm$ 11    & 0.160 $\pm$ 0.011 &  6.1 & 0.97 \\
           &  314.8 $\pm$ 3.5 & 0.173 $\pm$ 0.011 &  2.9 & 0.42 \\
           &  443.5 $\pm$ 4.2 & 0.144 $\pm$ 0.010 &  3.4 & 0.60 \\
           &  1483 $\pm$ ??   & 0.128 $\pm$ 0.010 & 10~~ & 2.00 \\
           &  234.4 $\pm$ 1.7 & 0.100 $\pm$ 0.010 &  1.2 & 0.32 \\
           &  153.4 $\pm$ 0.4 & 0.065 $\pm$ 0.010 &  0.5 & 0.21 \\
           &  109.6 $\pm$ 0.2 & 0.064 $\pm$ 0.009 &  0.4 & 0.15 \\
\enddata
\end{deluxetable}

\clearpage


\begin{figure}
\begin{center}
\epsscale{0.60}
\plotone{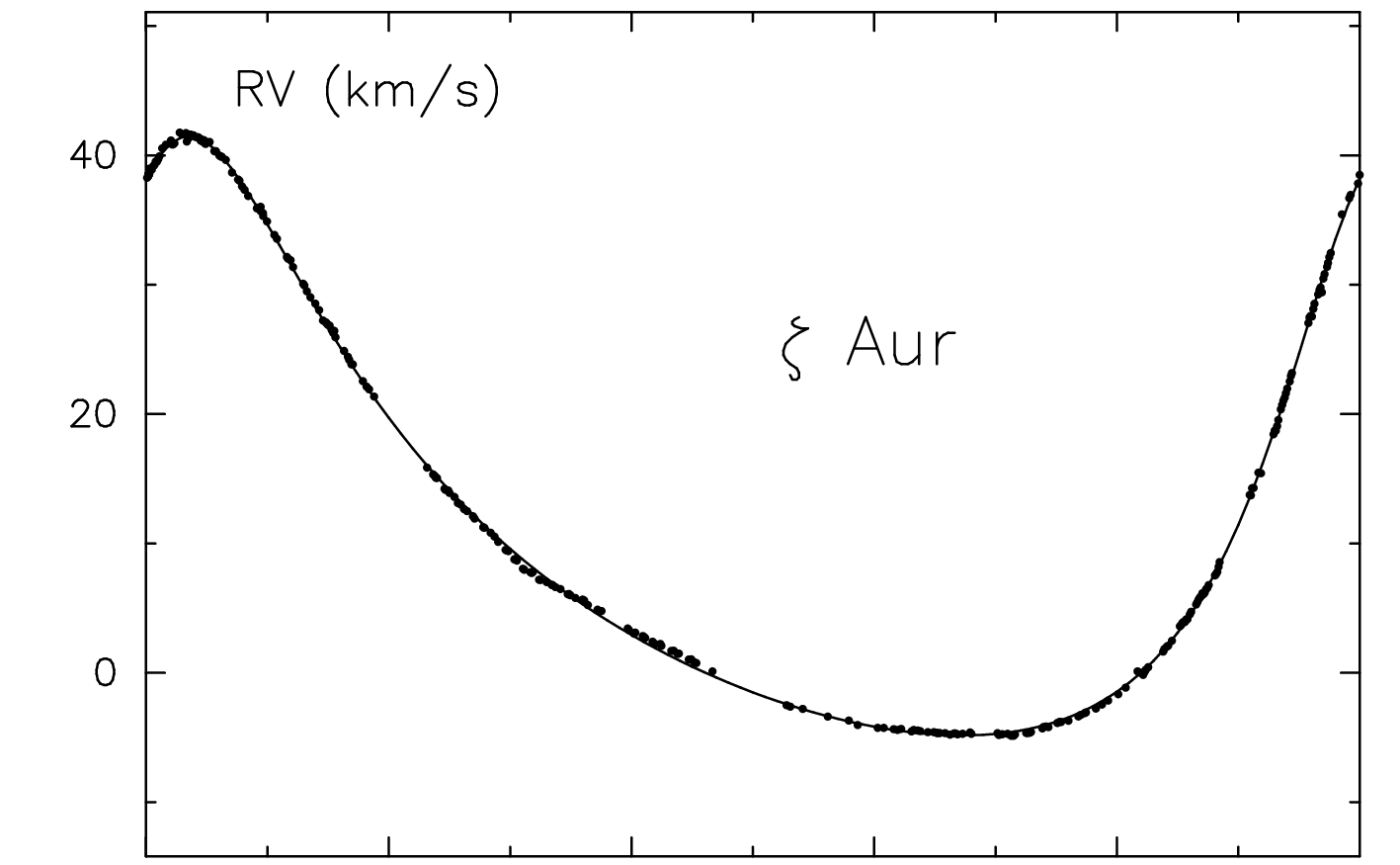}
\plotone{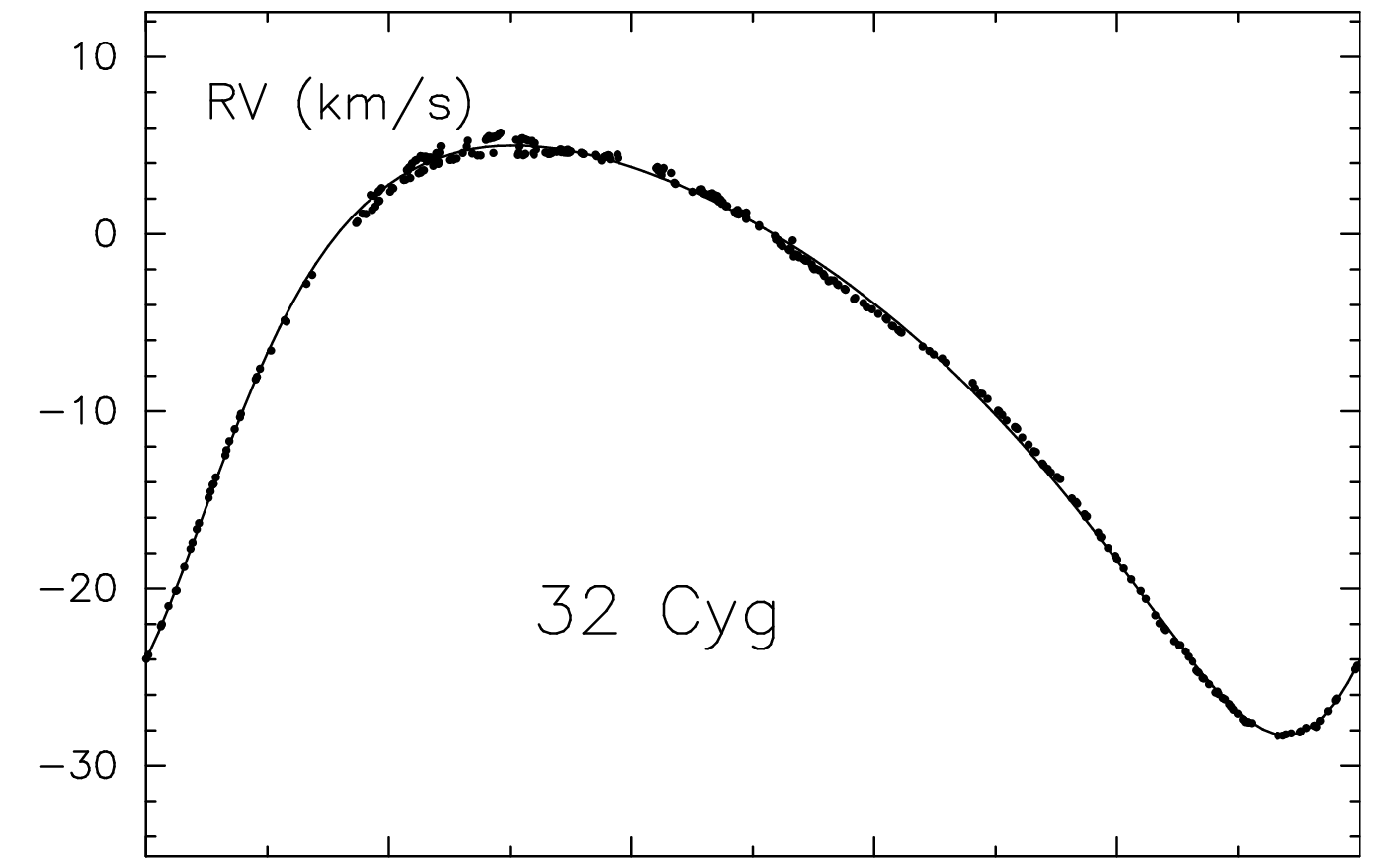}
\plotone{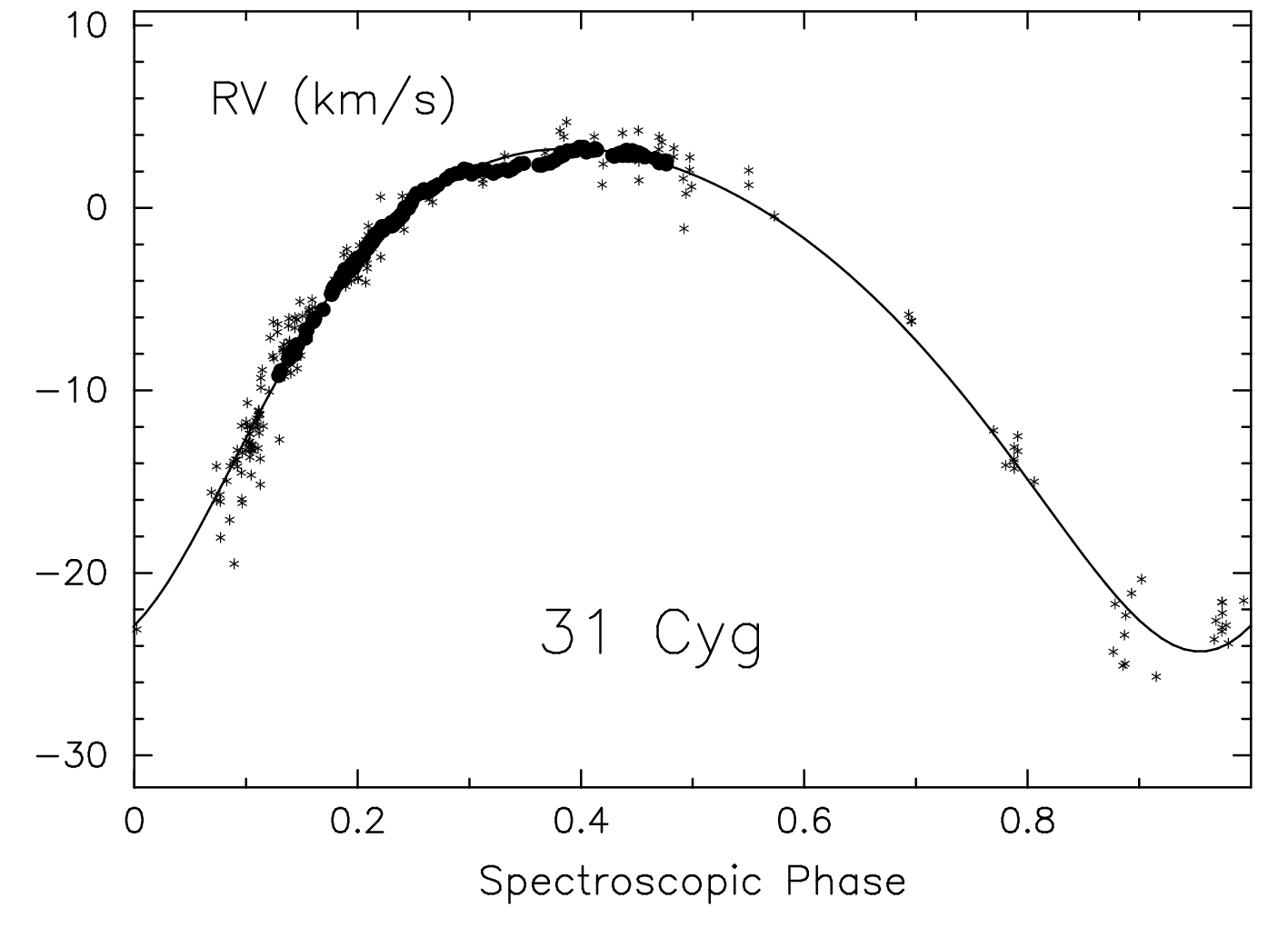}
\end{center}
\caption{Velocity curves of $\zeta$~Aur binaries.  Velocities from the AST are 
the dots.  The curves represent the orbital elements in Table 4.  The asterisks 
for 31~Cyg are the velocities listed by Wright \& Huffman (1968).
\label{fig1}}
\end{figure}


\begin{figure}
\begin{center}
\epsscale{0.60}
\plotone{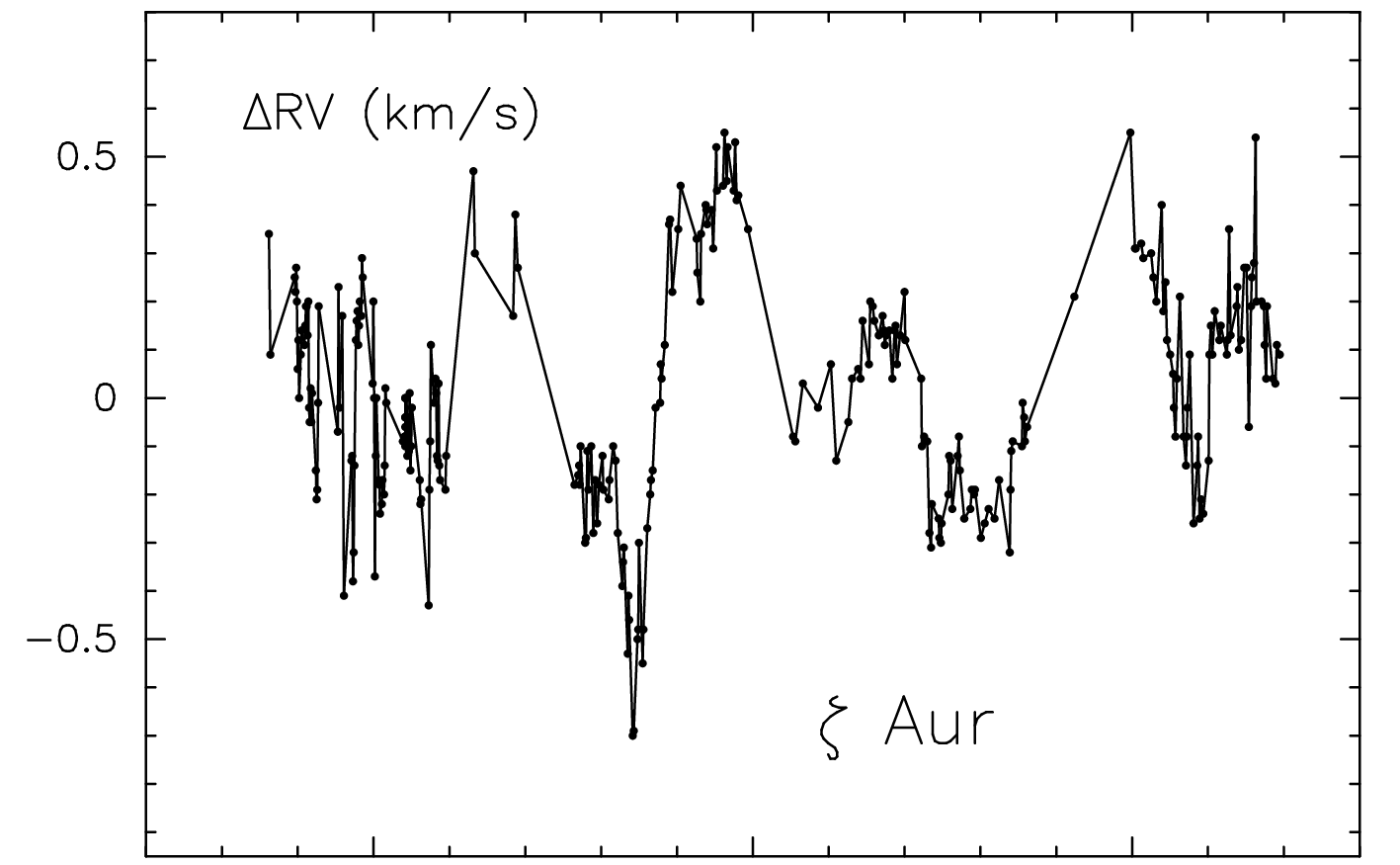}
\plotone{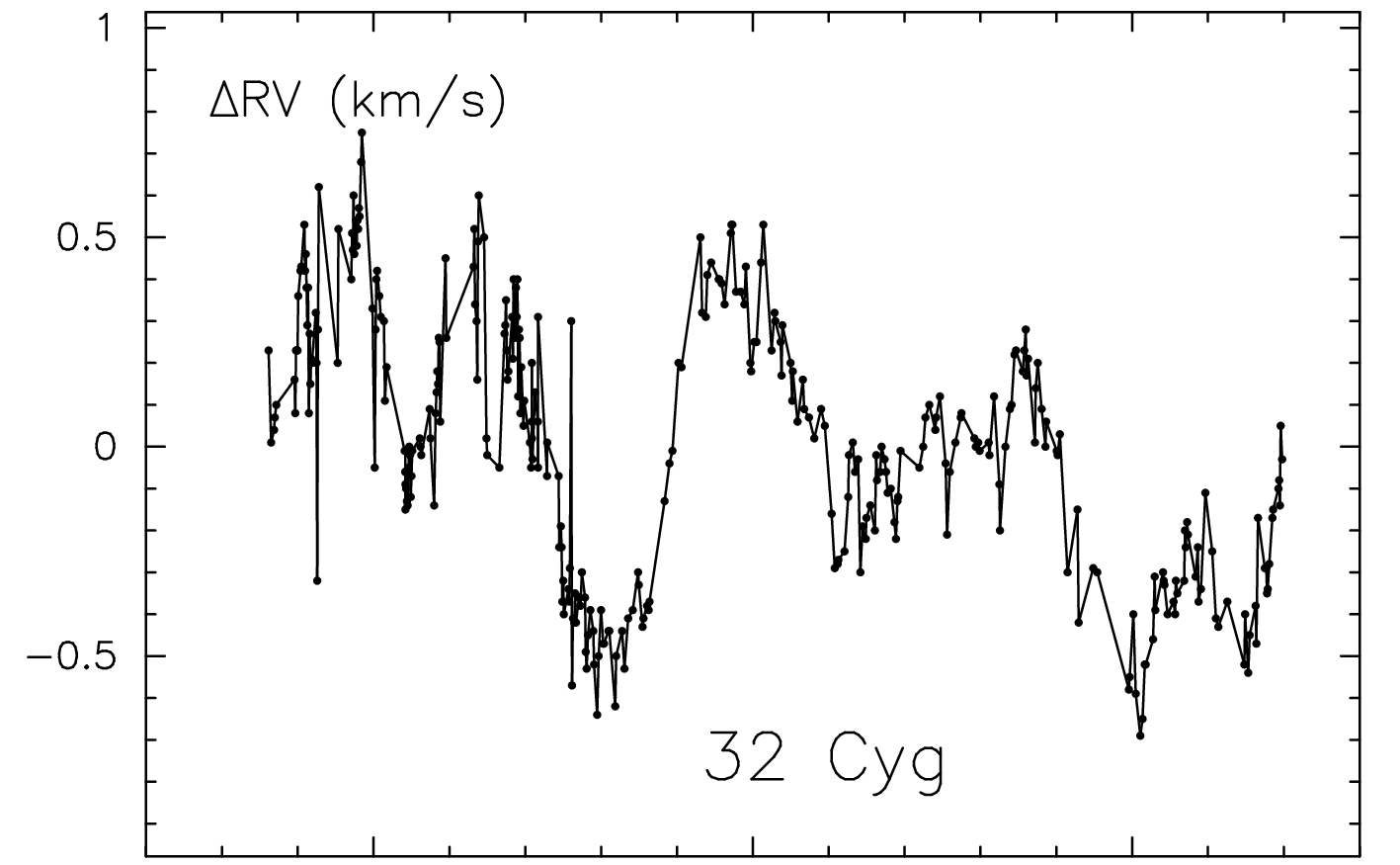}
\plotone{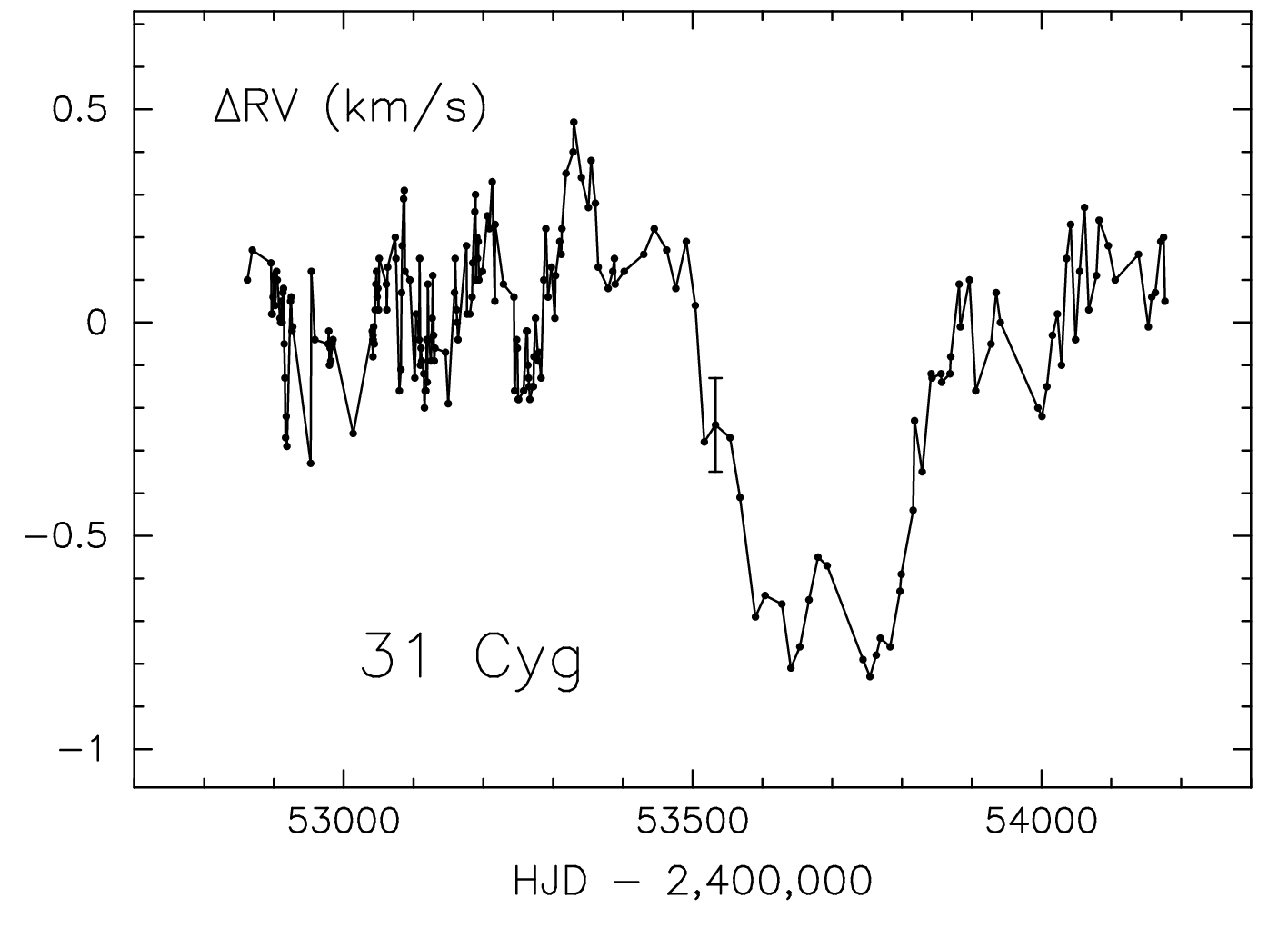}
\end{center}
\caption{Deviations from the fitted elliptical orbits.  Data from the AST are 
shown as dots.  We have connected the successive data points with line segments 
because we think this makes it easier to see and assess the shorter-term 
variability.  The errorbar plotted in the middle of the data string for 31~Cyg 
shows the level of uncertainty expected for all these data.
\label{fig2}}
\end{figure}


\begin{figure}
\begin{center}
\epsscale{0.50}
\plotone{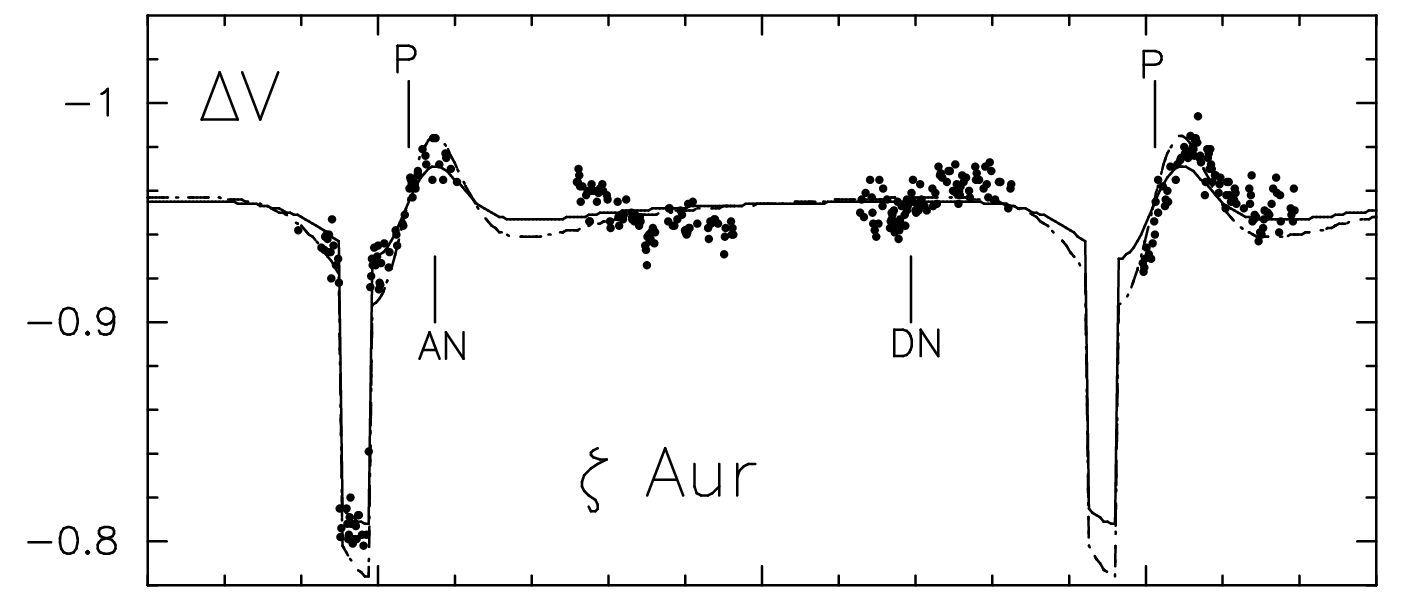}
\plotone{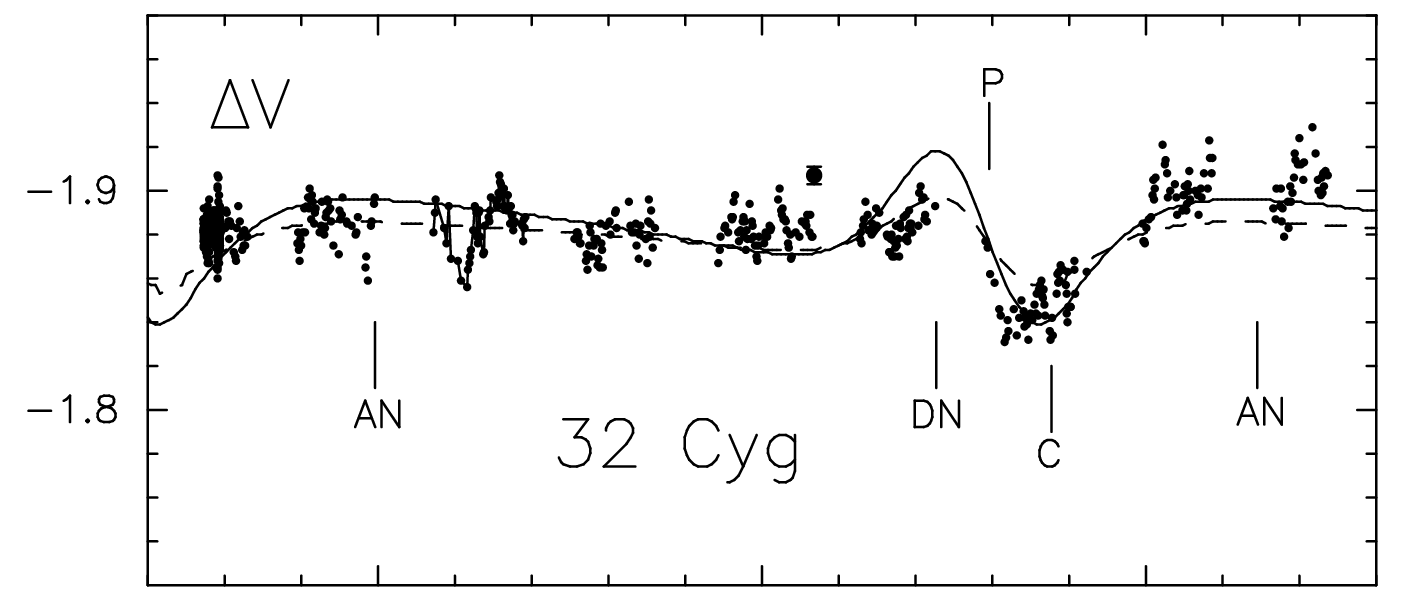}
\plotone{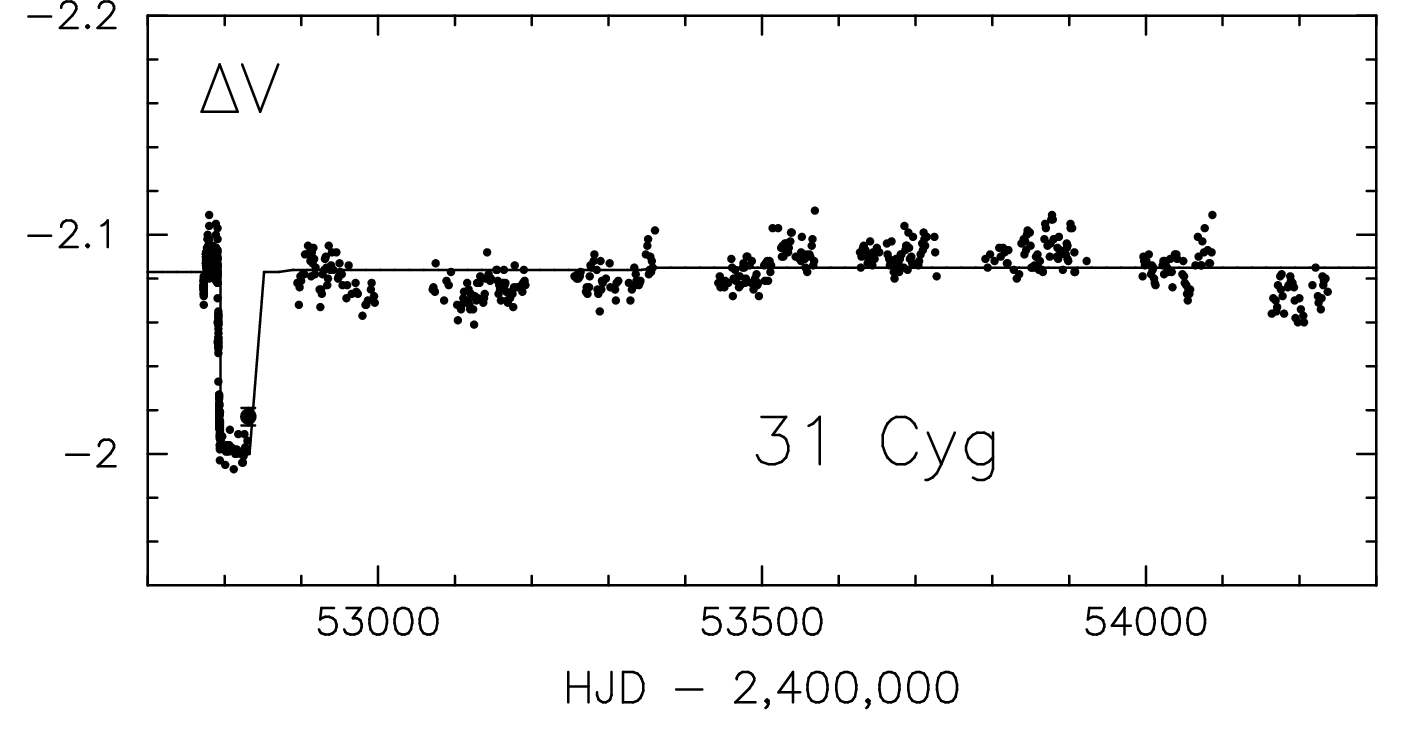}
\end{center}
\caption{Light variation of the three binaries.  These data for the $V$ band  
show two types of intrinsic variation, the normally intermittent variation from 
pulsation and ellipsoidal variation, to at least some extent.  The solid 
vertical lines mark times of periastron passage (``P") and nodal passages 
(``AN" for ascending node and ``DN" for descending node). The vertical line 
for 32~Cyg marked ``C" shows the time of conjunction (primary eclipse).  Thirty-one 
Cyg had a time of periastron passage at 52373, outside the range plotted.  The plotted 
curves represent the ellipsoidal variation calculated with the Wilson--Devinney code 
for these three systems.  The ellipsoidal variation for $\zeta$~Aur is quite prominent, 
peaking roughly a tenth phase after primary eclipse, as expected from the star's 
orbital orientation.  For 32~Cyg it contributes the prominent broad dip just before 
conjunction (primary eclipse).  The rather sharp eclipse is lost in the noise of these 
data but is obvious at $B$.  Pulsational variation is much harder to see in these data. 
There may be an example of it in the data around 53125, which we have connected with 
line segments.  The period would be 105 d.
\label{fig3}}
\end{figure}


\begin{figure}
\begin{center}
\epsscale{0.50}
\plotone{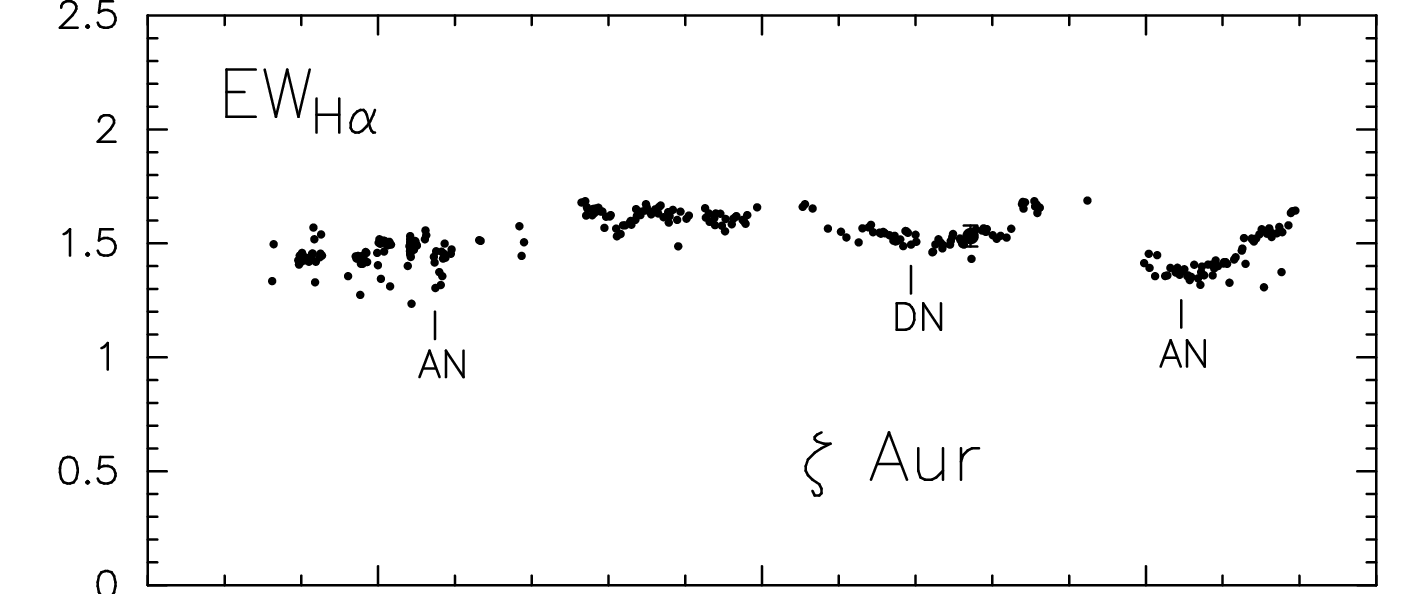}
\plotone{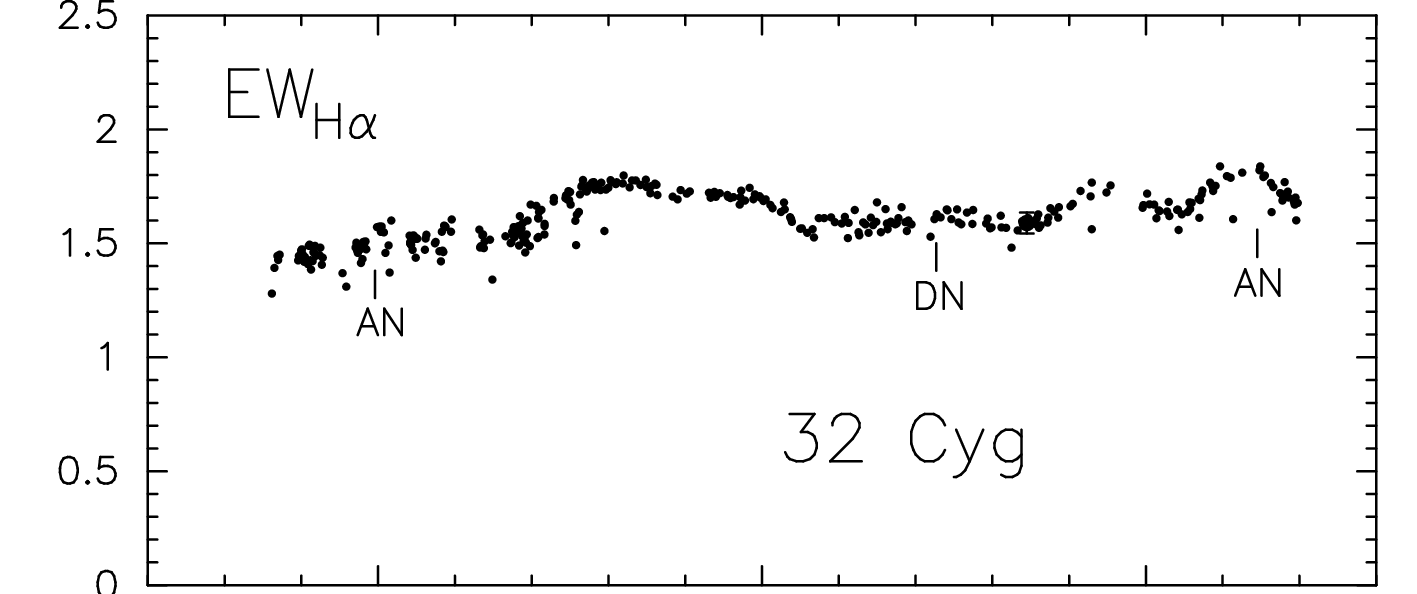}
\plotone{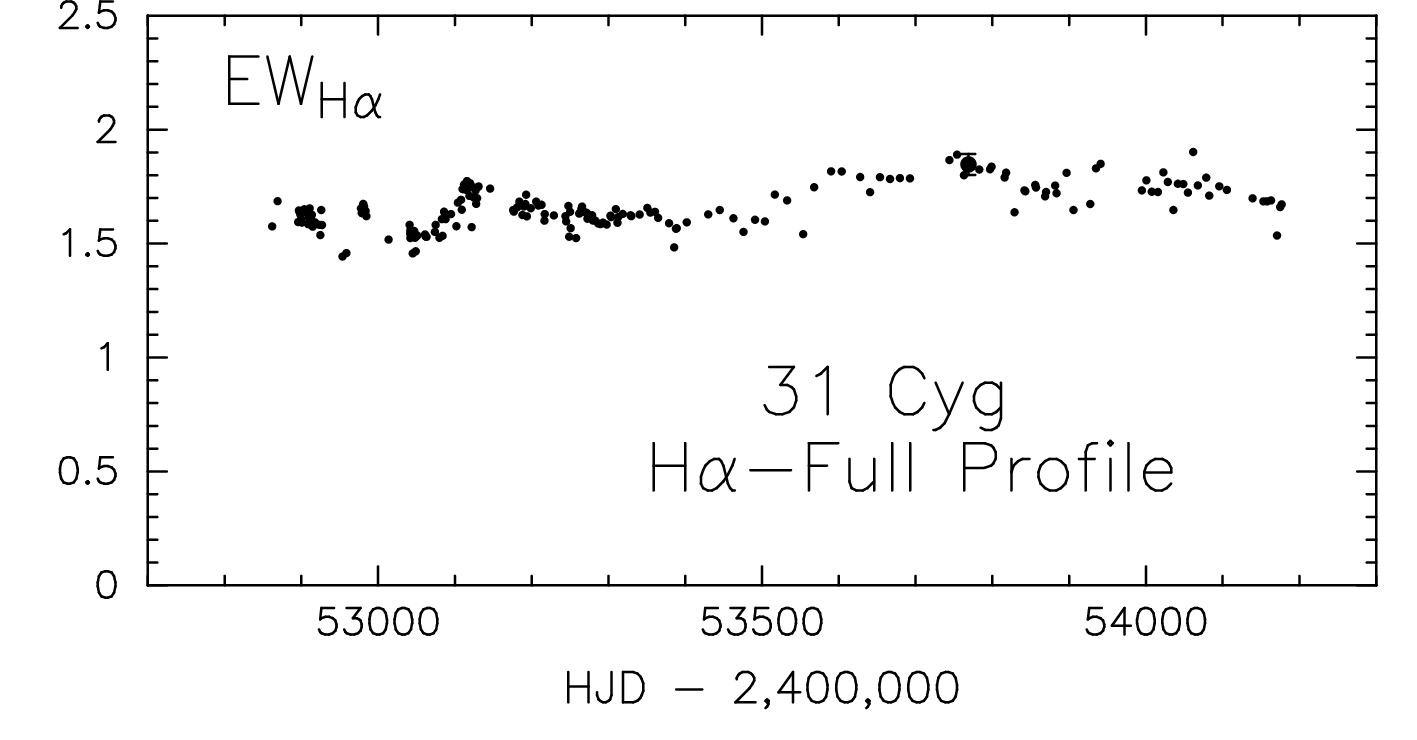}
\end{center}
\caption{Variation of H$\alpha$ in the three binaries.  This is the equivalent 
width of the bulk of the profile, EW1 in \AA, which shows how the chromospheric 
absorption changes globally.  Superior conjunction of the cool star, with its 
irradiated face in full view, occurs at 53245 for $\zeta$~Aur and 53475  
for 32~Cyg. 
\label{fig4}}
\end{figure}


\begin{figure}
\begin{center}
\epsscale{0.50}
\plotone{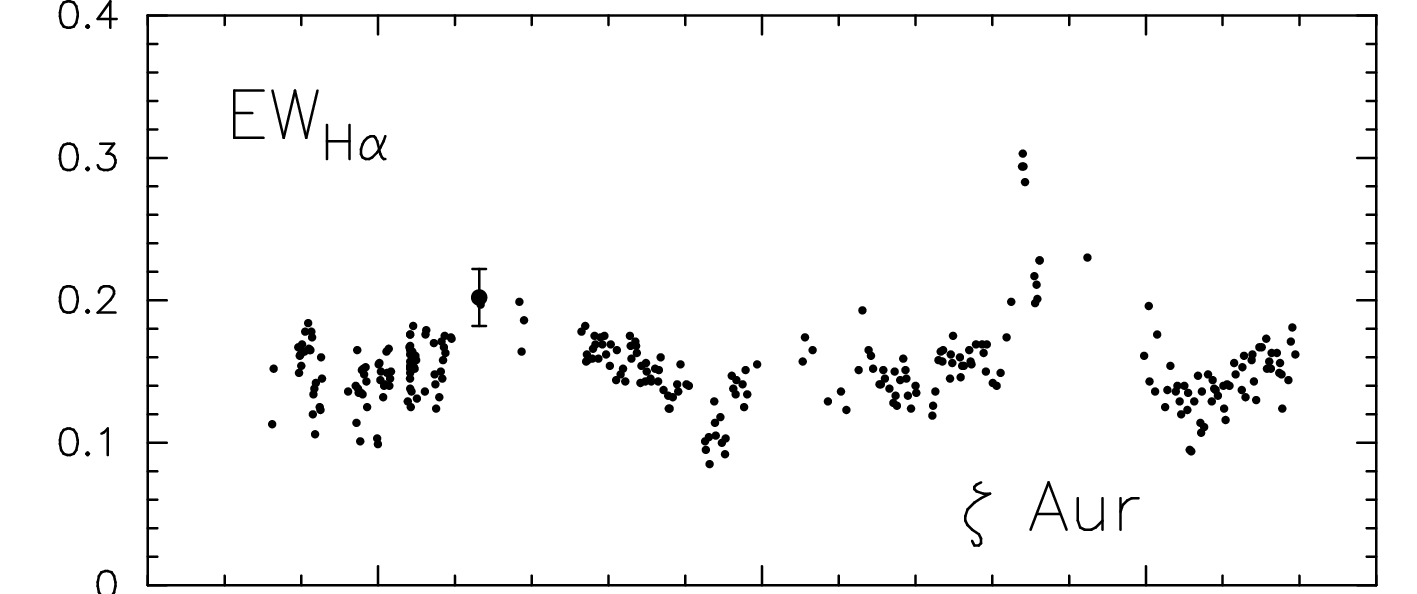}
\plotone{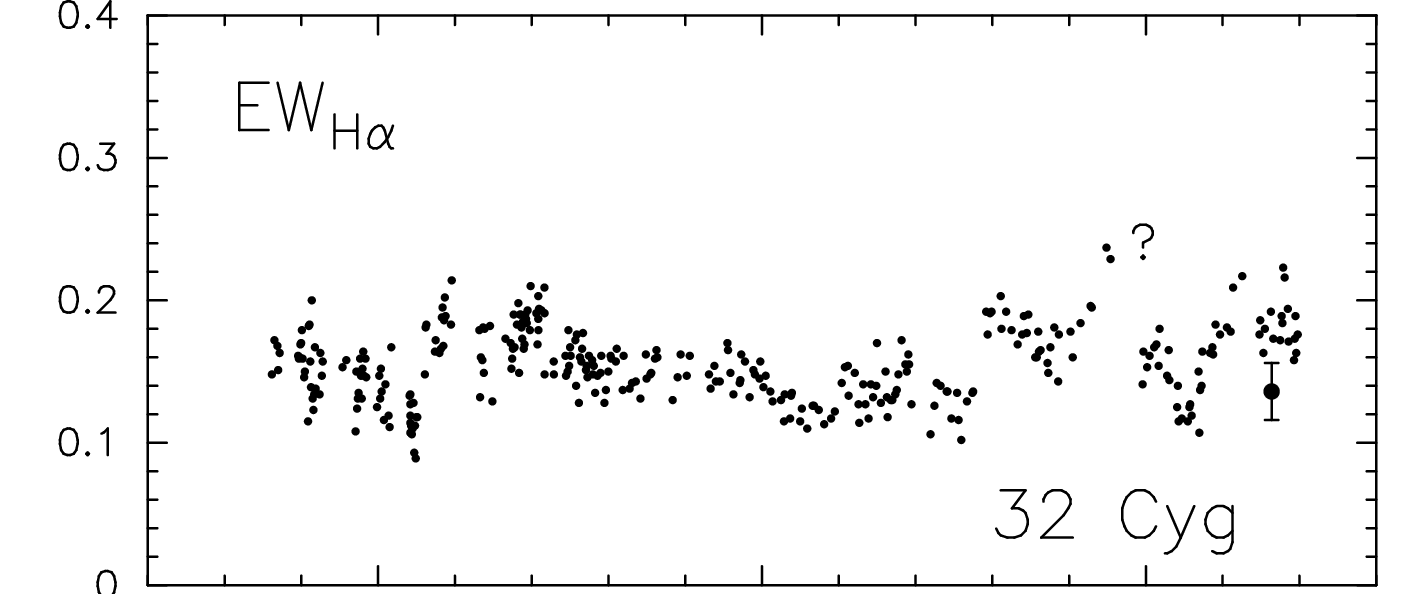}
\plotone{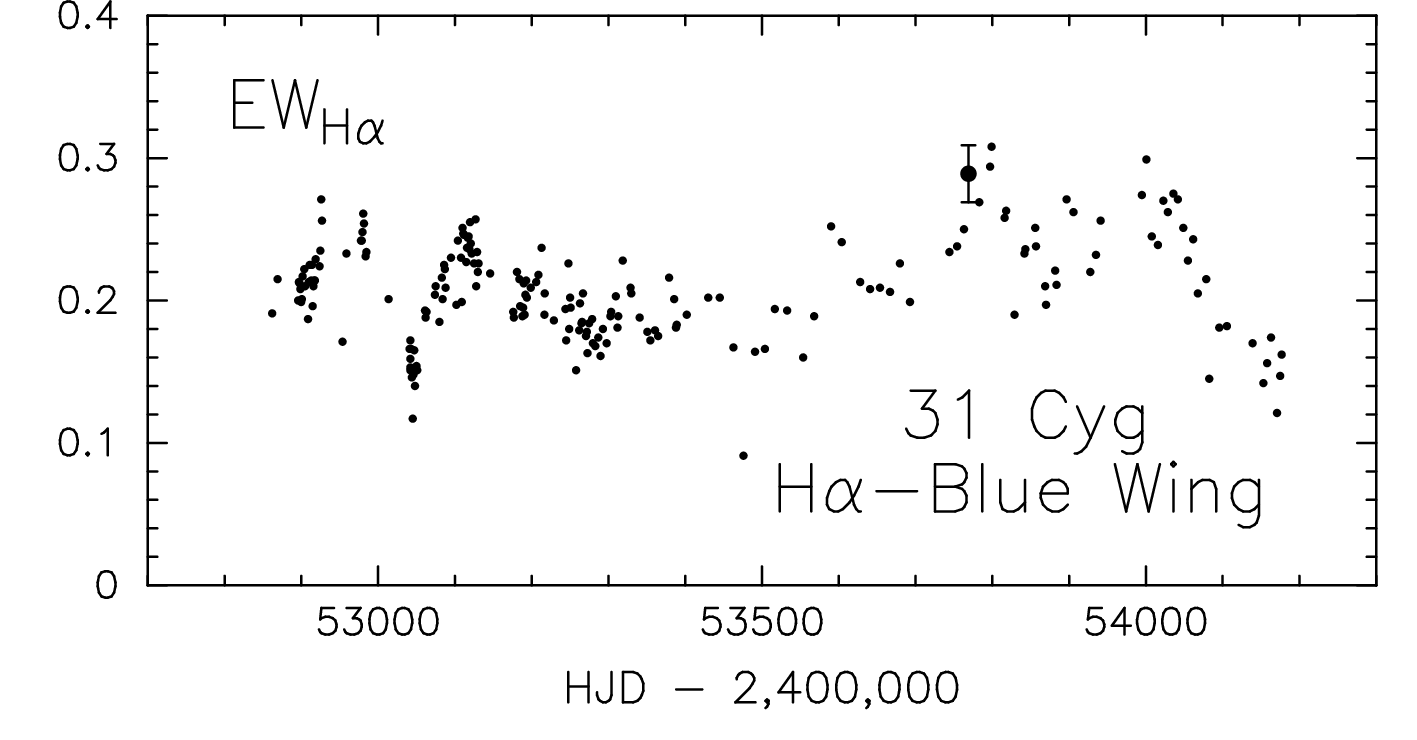}
\end{center}
\caption{Variation of the blue wing of H$\alpha$ in the three binaries.  
This is EW2 (in \AA), the band that measures enhanced wind absorption
in the blue wing of the profile beyond the velocity range of normal 
chromospheric absorptions.  Such enhancements are amply documented 
in a number of cool supergiants, but there are few of them detected 
in these data.
\label{fig5}}
\end{figure}

\end{document}